\documentclass[letterpaper]{article} 
\usepackage[]{format/aaai23}

\usepackage{times}  
\usepackage{helvet}  
\usepackage{courier}  
\usepackage[hyphens]{url}  
\usepackage{graphicx} 
\usepackage{natbib}  
\usepackage{caption} 
\usepackage{algorithm}
\usepackage{algorithmic}
\usepackage{hyperref}

\usepackage{xcolor}

\begin{document}


\title{Advances in Automatically Rating the Trustworthiness of Text Processing Services}




\author {
    Biplav Srivastava
    \textsuperscript{\rm 1},
    Kausik Lakkaraju
    \textsuperscript{\rm 1},  Mariana Bernagozzi \textsuperscript{\rm 2}, 
    Marco Valtorta
    \textsuperscript{\rm 1}
}
\affiliations {
    \textsuperscript{\rm 1} AI Institute, University of South Carolina, Columbia, South Carolina, USA\\
    \textsuperscript{\rm 2} IBM Research, Yorktown Heights, New York, USA \\

 biplav.s@sc.edu
}


\maketitle


\begin{abstract}

AI services are known to have unstable behavior when subjected to changes in data, models or 
users. Such behaviors, whether triggered by omission or commission, lead to trust issues 
when AI works with humans. The current approach of assessing AI services in a black box 
setting, where the consumer does not have access to the AI’s source code or training data,
is limited. The consumer has to rely on the AI developer’s documentation and trust that the system has been built as stated. Further, if the AI consumer reuses the service to build other
services which they sell to their customers, the consumer is at the risk of the service providers 
(both data and model providers). 
Our approach, in this context, is inspired by the success of nutritional labeling in 
food industry to promote health and seeks to  assess and rate AI services for trust from the perspective of an independent stakeholder. The ratings become a means to communicate the behavior of AI systems so that the consumer is informed about the risks and can make an informed decision. 
In this paper, we will first describe recent progress in developing rating methods for  text-based machine translator AI services that have been found promising with user studies. Then, we will outline challenges and vision for a principled, multi-modal, causality-based rating methodologies and its implication for decision-support in real-world scenarios like health and food recommendation.



\end{abstract}


\section{Introduction}

    

It is common to have real-world Artificial  Intelligence (AI) systems today that work with multimodal data like text, audio and image. Some illustrative examples are recommender systems (for food, movies),  collaborative assistants (for health, bank transactions and travel) and route finders (for traffic navigation, emergency evacuation). Here, sample AI services are - for natural language processing (NLP): entity detection, text translation, summarization and sentiment detection;  for audio - speech recognition, speaker identification; for image tasks - object detection, face identification, object counting; common tasks - reasoning, search and ranking. However,  the enthusiasm for AI services is also being dampened by  growing concerns about their reliability and trustworthiness with issues like bias (lack of fairness), opaqueness (lack of transparency), privacy of user data and brittleness (lack of robust competence) regardless of the data they use - be it text \cite{prob-bias-text}, audio \cite{prob-bias-sound} or image \cite{prob-bias-image}.  In fact, this has been argued as one of the key factors hampering the adoption of  AI techniques during an emergency like COVID-19 \cite{apollo-chatbots} or mainstream treatment like breast cancer \cite{breastcancer-health-ai-uk}. In this context, we explore how the trustworthiness properties of AI could be assessed and communicated, without access to its code or data (i.e., the black box setting), to promote growth.

 
\begin{figure}[h]
 \centering
   \includegraphics[width=0.4\textwidth]{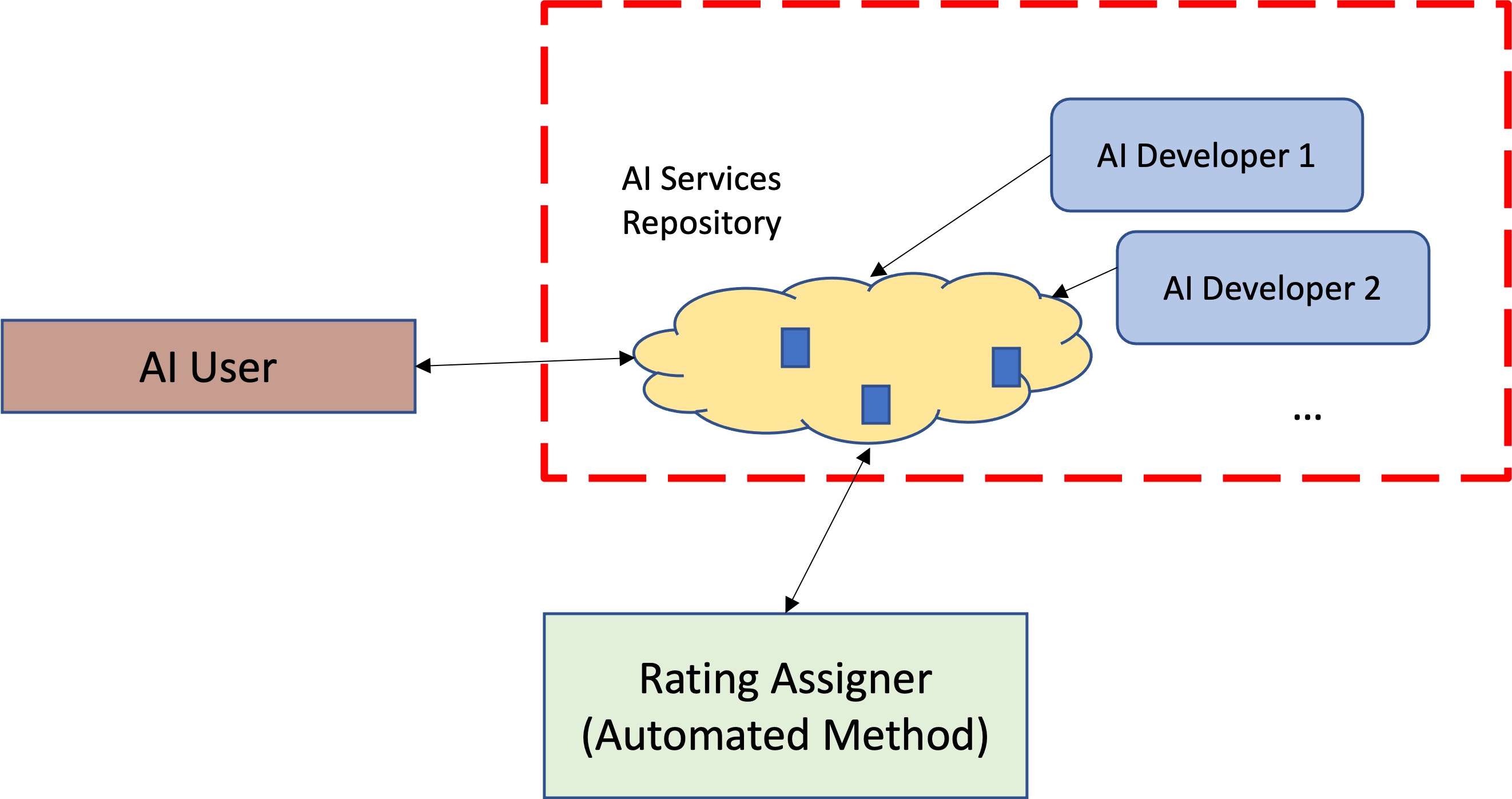}  
   \caption{The black box setting for AI services}
  \label{fig:bb-setting}
 \end{figure}

The black box setting for AI services is conceptually shown in Figure~\ref{fig:bb-setting}. Here, there is a repository which contains information to access AI services. Developers register their services as application programming interfaces (APIs). A user can search the repository, try out their capabilities, provide payment information and access services for production use based on their needs. Note that the user may be an end-user or even a developer building a new AI by reusing some of the capabilities of other AI or data services, thus creating {\em AI and data supply chains}. In such a setting, the user can only provide inputs but not access an AI's code or training data. An illustrative example of an independent API directory is ProgrammableWeb \cite{api-dir-pw} where multiple alternatives are available for the same capability (called {\em categories}) or Universal Description, Discovery, and Integration (UDDI) directories for web services \cite{uddi,ws-comp-survey}. The rating information from our proposed AI rating methods would help a consumer select the most suitable service for their requirements. Note that cloud services from commercial vendors (like Google, Microsoft and Amazon) restrict choice of services for a capability to usually one forcing the consumer to switch platforms if their needs are not met or bear with inadequate capabilities if the consumer needed multiple capabilities of which only some are fulfilled. An independent API directory promotes consumer choice and competition among service (API) providers.

Our idea here is to invoke AI services in a controlled evaluation setup with specific input data and observe the output. The results are then statistically assessed and labels are assigned. These label,  also called {\em ratings}, which communicate an AI's behavior under test conditions, can be attached to its information in a services repository  and will be open to public review. We argue that much like nutrition labels for  packaged foods, and safety and fuel economy ratings for cars, such ratings for AI behavior would help consumers select AI services based on their application needs. 

In the rest of the paper, we first  provide background on the problem of AI trust and the {\em transparency through documentation}  approach of building trust. A good example of this is  nutrition labeling for prepared foods in the United States which we use as a motivation.  
We then summarize results from our initial approaches for rating machine translators and chatbots that were found promising. Then, we articulate a more general path forward grounded in a causal setting and discuss 
several research questions along the way.

\section{The Problem of AI Trust}

Trust is a multi-faceted quality that is studied in the context of humans in humanities, and now increasingly gaining importance in AI as systems and humans collaborate closely. In \cite{trust-ml-book}, the author identifies  a trustworthy machine learning system as one that can demonstrate four desirable properties:  sufficient competence (performance), reliability, support for human interaction and alignment with human purpose.  Among them, while competence (first) and reliability (second) may be obvious since they are also applicable for other engineering systems, support for human interaction  (third) implies that the system be explainable and alignment with  human values (fourth) suggests the system to have qualities like fairness, ethics and morality. There are well-studied approaches to measure the first two and growing literature in the third 
\cite{trust-explainablemetric-survey} to measure explainability.
Hence, we will focus on the fourth category noting that any rating has to cover multiple dimensions.

Humans values are motivated by the fact that humans are social agents who live in a community, and
ethics and morality are ways to guide our behavior so that both
social and individual wellbeing is coherently achieved and maintained.
Therefore, we usually constrain our decisions according to
moral or ethical values that are suitable for the scenario in which
we live. 
While ethical principles are not universal and can vary according
to scenarios, tasks, domains, and cultures, one property that is
usually included in the realm of ethical behavior is fairness. It is 
the impartial and just treatment or behavior without favoritism
or discrimination fairness studied both in AI and philosophy \cite{bias-survey}. The absence of fairness is  referred to
as bias.

From an engineering perspective, trust is an important non-functional requirement  in AI development, deployment,
and usage. Users and stakeholders should be able to
have a justified trust in the AI systems they use, otherwise
they will not adopt them in their everyday life. 
In the same way as humans, AI systems that have an impact on real
life environments or on humans, or that recommend decisions to be
made by humans, should be designed and developed in a way that
they follow suitable ethical principles as well. A line of work in AI looks at how to teach an AI system to act within ethical or legal guidelines \cite{book-moral-machines}.

We will call trust dimensions in human values as {\em issues}. For rating methods to be implemented, we only require that   automated {\em issue checkers}  should exist to measure the issue on a a given AI. 
Previously, we had used the issue checkers of abusive language (AL) \cite{hateoffensive}, bias (B)   \cite{Hutto2015ComputationallyDA}, information leakage (IL) \cite{ethical-dialog}, complexity checker implemented by \cite{dialog-complexity} (CC)  in our work on rating chatbots  \cite{chat-rating}.  We will focus on bias.

Although fairness has been studied extensively in the AI literature for both bias detection and mitigation, there is no end in sight on how to achieve it satisfactorily.
As per a  recent taxonomy of bias in a AI system pipeline \cite{bias-ai-practice},  a system can exhibit bias during data creation, problem formulation, data analysis, and validation and testing biases. 
To detect bias, one needs to  select a measure of fairness from the many proposed in literature \cite{bias-survey,bias-definitions-eg}.
They can be broadly grouped into the following categories: (a) {\em individual fairness} definitions which  measure whether similar individuals are treated similarly, (b) {\em group fairness} definitions where statistical properties on sub-groups of the population are measured, and (c) {\em hybrid} where a combination is measured. 
Example of individual fairness is {\em causal discrimination} which requires a model to produce the same output for any two instances that have the exact same attributes and  example of group fairness is  {\em equalized  odds} which requires false positive and negative rates to be similar amongst protected and non-protected groups.
However, it is well known that definitions prevalent in  the machine learning literature are not what  are legally enforceable \cite{bias-definitions-legal}, and hence, this continues to be an ongoing research area.

\subsection{Trust in Text-based AI Services}

For space reasons, we now narrow the discussion to text-based AI but similar issues have been found for AI working with other data formats. 

\subsubsection{Bias in Machine Translators}

Machine translators are a popular form of AI used widely in applications.
In \cite{translator-assess-bias}, the authors test {\em Google Translate} on sentences like  "He/She is an Engineer" where occupation phrase ({\em Engineer} in example) is from those recognized by the U.S. Bureau of Labor Statistics (BLS) \cite{occupation-list}. They compare frequency of female, male and gender-neutral pronouns in the translated output  with BLS data about expected frequency.
In another paper \cite{translator-bias}, the authors look at a transformer architecture for machine translation, Open NMT translator\footnote{At http://opennmt.net/} and two debiasing word embeddings. They consider sentences of the form: “I’ve known \{her/him\} $\prec$proper noun$\succ$ for a long time, my  friend works as \{a/ an\} $\prec$occupation$\succ$.”
They consider translation from English to Spanish and look at the linguistic form of the noun phrase used for {\em friend}  based on {\em occupation}. They make a list of 1019 occupations publicly available\footnote{At: https://github.com/joelescudefont/genbiasmt}. In \cite{trans-rating,trans-rating-jour}, we consider the translation setting from English to an intermediate language, and then from it back to English. Inputs are primed with gendered pronouns and checked for variation. In all the settings, significance differences are found which is evidence of bias behavior based on {\em group fairness} definitions.

\subsubsection{Bias in Sentiment Assessment Systems}

Another popular form of AI  is a sentiment analysis system (SAS) which given a piece of text, assigns a score conveying the sentiment and emotional intensity expressed by it. 
In \cite{sentiment-bias}, the authors experimented with sentiment analysis systems that participated in SemEval-2018 competition. They created the Equity Evaluation Corpus (EEC) dataset which consists of 8,640 English sentences where one can switch a person’s gender or choose proper names typical of  people  from different races. They also experimented with Twitter datasets. The authors found that up to 75\% of the sentiment systems can show variations in sentiment scores, which can be perceived as bias based on gender or race.

While much of the work in sentiments has happened for English, there is growing interest in other languages. In  \cite{sentiment-multilingual}, the authors re-implement sentiment methods from literature in multiple languages and reported accuracy lower than published. Since multilingual SASs often use machine translators which anyway can be biased, and further acquiring training data in non-English languages is an additional challenge, we hypothesize that multilingual  SASs could exhibit  gender bias in their behavior.

\subsubsection{Adversarial Attacks Using Sentiments}

The instability of SAS is often a source of adversarial attacks in AI systems. This can manifest as adversarial examples \cite{adv-example}, which modify test
inputs into a model to cause it to produce incorrect outputs (sentiments for SAS), or 
backdoor attacks \cite{backdoor} which compromise the model by poisoning the training data  and/or modifying the training. A more sophisticated form of attack is proposed in
\cite{spin-attack} where a high-dimensional AI  model, like for  summarization, outputs positive summaries of any text that
mentions the name of some “meta-backdoor" like individual or organization.

\subsubsection{Handling Bias}

There have been prior efforts to address bias by improving handling of training data, improved methods or post-processing on the AI system's output \cite{bias-survey}.
However, the research in bias in NLP systems has shortcomings. In \cite{prob-bias-text}, the authors report that quantitative techniques for measuring or mitigating "bias" are poorly matched to their motivations and argue for the need to conduct studies in the social context of real world. 
Furthermore, work is needed to align technical definitions  with what is legally enforceable \cite{bias-definitions-legal}. 


There is also increased interest in exploring causal effects for AI systems. For natural language processing (NLP),   \cite{causal-nlp} provides a survey on estimating causal effects on text. Similarly, causal reasoning is being used for object recognition systems \cite{mao2021generative} and recommendation systems  \cite{causal-collab}. But they do not consider using such analyses for communicating trust issues.

We now look at the new idea of assessing and rating AI services from a third party perspective in a black-box setting  which we discuss next.

\section{Transparency Through Documentation}

To promote user trust in a product or service (collectively called an offering), transparency is recognized as  a promising technique. It can be created  by documenting either the process of creating and delivering the offering or documenting characteristics of its output. It can be documented by the   provider of the offering, consumer or an independent third party. For example, ratings for fuel economy of cars and energy consumption of electrical devices  are created by manufacturers while the Software Engineering Institute (SEI) Capability Maturity Model certification of a software organisation is produced for the process of software development by an independent organization.
In AI, datasheets has  been introduced as a form of documentation to promote transparency \cite{datasheets}.

\subsection{Lessons From a Workable Solution - Nutrition Labeling
for Prepared Foods}

Most prepared foods in the United States are required to carry a label conveying its nutritional information \cite{food-label-fda,food-label-ucd}. They are primarily regulated by the US Food and Drug Administration (FDA); other agencies may regulate too based on food types like meats and poultry by  the United States Department of Agriculture (USDA) Food Safety and Inspection Service (FSIS). The templates for the label are created by the regulator along with guidelines for how to assess food for their nutritional value, but the food manufacturer places the labels on the food. 

Such labels have been found to be helpful to consumers. For example, one study found that {\em "76 percent of adults read the label when purchasing packaged foods, and that more than 60 percent of consumers use the information about sugar that the label provides"} \cite{food-label-report-ucs}. It is a different matter that nutrition labeling is a work-in-progress with requirements evolving and  the industry sometimes trying to pushback and dilute the labeling requirements.  

The AI trust labels we envisage will similarly help the users make informed decisions while selecting and using AI. However, they would be different from food labels in that the labels would be created by independent third-parties, i.e., who are neither the AI developers nor consumers. They are created by subjecting AI to specific inputs, analysing the outputs in the context of inputs given, and assigning labels. We now consider two specific efforts in this direction - for machine translators and chatbots, respectively.

\subsection{Initial Approach - Rating Machine Translators}

In \cite{trans-rating-jour,trans-rating}, we proposed a technique to rate automated machine language translators for gender bias. Further, we created visualizations to communicate ratings \cite{vega-rating-viz} to users, and conducted user studies to determine how users perceive trust in the presence of such ratings \cite{vega-user-study}.

\begin{figure}[h]
 \centering
   \includegraphics[width=0.3\textwidth]{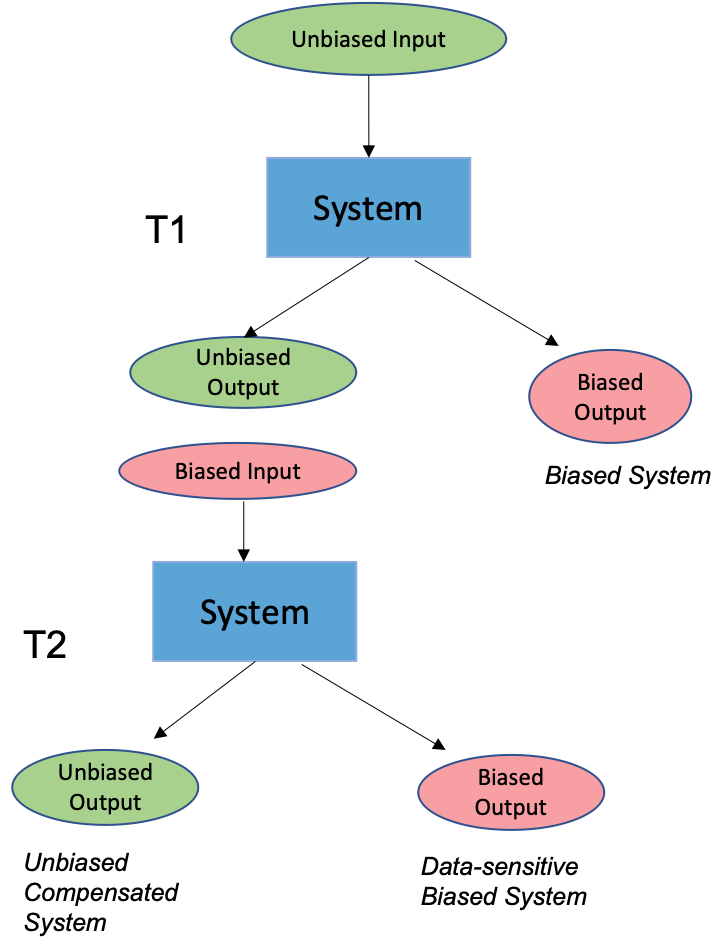}  
   \caption{A 2-step approach to rate an AI service from \cite{trans-rating-jour}.}
  \label{fig:rating-tree}
 \end{figure}

We now summarize the method which is illustrated in Figure \ref{fig:rating-tree}. First, unbiased and biased data is prepared for translation with respect to the number of gender variables. For this purpose, input is made up of two sentences containing
one gender place-holder each. Its format is: {\em
$\prec$Gender-pronoun$\succ$ is a $\prec$Occupation-Performer$\succ$. $\prec$Gender-pronoun$\succ$ is a $\prec$Occupation-Performer$\succ$}. We chose this two-sentence format so that the text  could include multiple gender values, and   expose in
a more articulate way the possible bias translation issues.
An example is - {\em She is
a Florist. He is a Gardener}.

In the first stage (T1), the AI system under consideration is given unbiased input and its output is analyzed. If the output is biased, the system is rated Biased (BS) under that test.  This means that the system introduces bias even when the input is unbiased. This is the worst rating that could come out of the procedure.

 
If on the contrary the output is unbiased, the system is now subjected to biased input (T2) and its output analyzed. If the output is biased, the system is rated Data-Sensitive Biased System (DSBS). This means that 
the system does not introduce bias but it follows whatever bias is present in the input. 
If instead the output is unbiased, given biased input in the testing stage T2, then the method would output the rating Unbiased Compensated System (UCS). 
This is the best rating and it means that the system not only does not introduce bias, but it does not even follow the bias of the input data,
and is instead able to compensate for possible bias in the input data.



We also introduced a method to rate
composite AI services of the form $A_i \star A_j$, where $\star$ is sequential composition.  The method is a decision table shown in Table~\ref{tab:seq-comp}. Here,
the predecessor service is shown in rows and the successor service is shown in column. 
So, if a biased system $A_i$ is sequentially composed with another biased system $A_j$, the outcome of the composite service can be biased (BS), unbiased compensation (UCS) or data-sensitive biased (DSBS). If the predecessor is DSBS, the rating of the composite will depend on the rating of the successor.

 
\begin{table}
\centering
    \begin{tabular}{|l|l|l|l|}
    \hline
          $A_i$ $\star$ $A_j$ & BS & UCS & DSBS  \\ \hline \hline
     BS   & BS/ UCS/ DSBS & UCS & BS  \\ \hline
     UCS    & BS & UCS & DSBS  \\ \hline
     DSBS    & BS & UCS &  DSBS \\ \hline                               
    \end{tabular}
    \caption{Sequential Composition of APIs as described in \cite{trans-rating-jour}. The labels of the rows are for the first service, while the labels of the columns are for the second service.}
    \label{tab:seq-comp}
\end{table}


\subsection{Initial Approach - Rating of Chatbots}

\begin{figure}[h]
 \centering
   \includegraphics[width=0.5\textwidth]{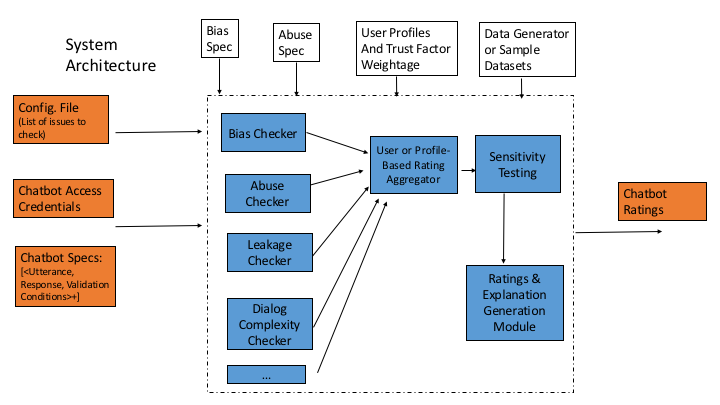}  
   \caption{Architecture of the chatbot rating system presented in \cite{chat-rating}}
  \label{fig:chat-rating-arch}
 \end{figure}

In \cite{chat-rating}, we introduced a new approach to rate dialog systems, i.e., chatbots, based on their behavior with respect to a list of configurable k issues, such as bias
(B), abusive language (AL), conversation complexity (CC),
and information leakage (IL), for all of which, checkers are available in literature. The system is conceptually illustrated
in Figure~\ref{fig:chat-rating-arch}. Its inputs are the issues to be considered,
the access details of the chatbot to be rated so that they can be invoked (online testing) or a dataset containing past conversation using the chatbot (offline testing) and user profiles with respect to which test results have to be assessed. User profiles capture relative weightage of issues from the perspective of a group of users like privacy-conscious users.
The output is a rating for
the chatbot, conveying its level of trustworthiness for a specific
user or user’s profile.

The system can  assign a type of rating, conveying
a: {\em Trustworthy agent (Type-1)}, which starts out trusted with
some score (Low, Medium or High) and remains so even after considering
all variants of models, data, and users; {\em Model-sensitive trustworthy
agent (Type-2)}, which can be swayed by the selection
of a model to exhibit a biased behavior while generating its
responses; {\em Data-sensitive trustworthy agent (Type-3)}, which
can be swayed by changing training data to exhibit a biased
behavior; {\em User-sensitive trustworthy agent (Type-4)}, which can
be swayed by interaction with (human) users over time to
exhibit a biased behavior; or {\em a Multi-sensitive agent (Type-N)}, which
can be swayed with a combination of factors. See \cite{chat-rating} for further details.

 
\section{Challenges in Rating  Multimodal AI Systems of the Future}


We now first look at significant AI applications that rely on users to trust them for success, and then draw challenges for rating methods.
We will call them {\em trust critical applications}.

The first area we consider is food.
Food is not only an innate psychological need for human life, it is also a key factor driving a society's health and economic well-being. Maintaining a healthy diet is essential for good health and nutrition as it prevents many chronic non-communicable diseases. The importance of diet and food has attracted increasing attention, leading to the coining of a new term called as \textit{precision nutrition} \cite{chatelan2019precision,wang2018precision} - which seeks to better the health of a person through precise dietary intake based on unique characteristics of an individual such as DNA, race, gender, and lifestyle habits. The process of cooking is a key enabler for precision nutrition, which has led to extensive research in building and deploying decision-support artificial intelligence (AI) tools - varying from information retrieval (IR) using interfaces to task-oriented chatbots \cite{min2019survey,jiang2020food,r3-food,food-reco-india-icdmw22}.

The second area we consider is using AI for disseminating official information about elections, and in particular among senior citizens \cite{senior-elections}. 
While senior citizens (individuals 65+) exhibit some of the highest levels of voter turnout \cite{election-2016-voting}, many potentially willing voters are unclear about disability accommodations at election sites \cite{senior-election-participation}, and some are uncertain about the basics of where and how to register and to vote \cite{senior-election-participation-2}.  While government entities have online resources to educate voters, senior citizens are still disproportionately prone to believe and share false information. A recent study suggested that seniors are seven times more likely to share articles from misleading or false websites than individuals under 29 \cite{fakenews-senior}.  
In this context, there is initial evidence that a chatbot could help drive voter participation: both anecdotally with official information about election processes \cite{safe-chatbot-election} as well as in an experiment during a 2019 Wisconsin statewide election with a simple reminder to vote and providing logistical details \cite{voter-chatbot-rct}.

\subsection{Research Challenges}
We now identify and discuss key research  challenges to make rating feasible for trust-critical applications like those identified above dealing with human living and governance.

\noindent {\bf Challenge 1: Data availability}. To assess an AI system, we need to invoke it using data that is independently obtained. Two promising directions of research here are  of creating synthetic data generators, like \cite{synthetic-sdv-datagen}, and sampling data from large data repositories like Imagenet \cite{imagenet} and Wikipedia. 
One could also use datasheets as a way to promote transparency  of data used during testing \cite{datasheets}.

\noindent {\bf Challenge 2:  Assessing trust behavior}. We assume that each trust issue has a corresponding issue checker available. However, identifying, implementing and improving checkers is an ongoing area of research. We referred to checkers for abusive language (AL) \cite{hateoffensive,twitter-curse}, bias (B)   \cite{Hutto2015ComputationallyDA}, information leakage (IL) \cite{ethical-dialog} and conversation complexity checker \cite{dialog-complexity} (CC). For NLP models,  \cite{sear-acl18} presented
a method to generate semantically equivalent test cases that
can flip the prediction of learning models and can serve as a principled checker for many issues.

\noindent {\bf Challenge 3: Assigning labels}.  Although it is desirable to create rating labels with clear and consistent semantics, it is an open question what these should be. For example, a rating that just differentiates between unbiased and biased behavior, leaving others as undecided, may leave too many cases undecided for it to be of practical use. In the two presented cases, the ratings for machine translators and chatbots are useful from the perspective of individual AIs, but do not generalize. It is also unclear whether the ratings should convey only
 statistically rigorous information or also  socially relevant information.

\noindent {\bf Challenge 4:  Handling composition}. Real world AI applications use a variety of data modalities and AIs. For a rating method to scale to large applications, the rating methods have to be compositional so that the rating for an aggregate AI could be computed from that of its parts rather than creating a large testing setup taking into account every combination of data and AI.  

\noindent {\bf Challenge 5:  Validating ratings with users perception}. Since the ratings created by automated methods are intended to help people make decisions about selecting and using AI, one should also evaluate them in user trials for effectiveness. In the presented case of machine translators \cite{vega-user-study,vega-rating-viz}, we created a visualization tool for rating  and conducted a detailed study  to find how users relate to bias in translators and whether bias ratings were useful (which they were). More such studies are needed to connect the social and technical considerations involved in using AI. 


\section{Towards a Principled Multimodal Rating Approach in a Causal Setup}

\begin{figure}[h]
 \centering
   \includegraphics[height=0.20\textwidth]{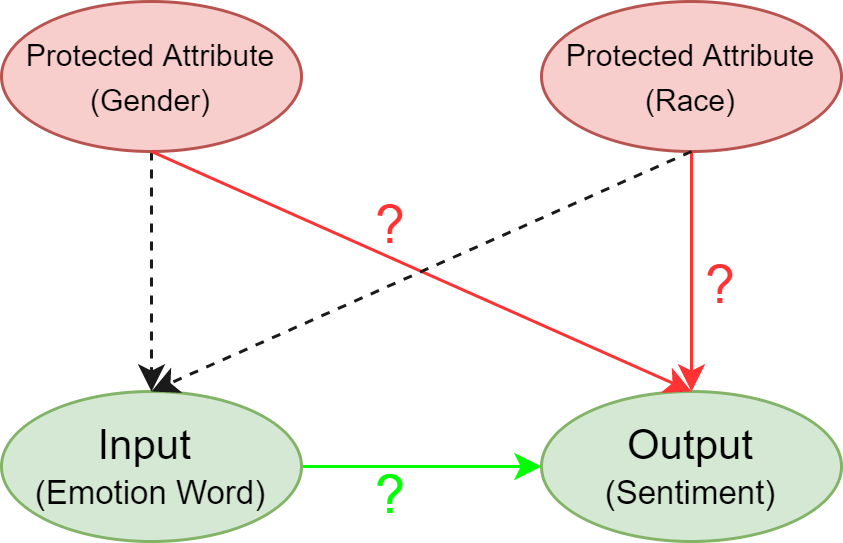}  
   \caption{Our proposed causal model}
  \label{fig:causal-model}
 \end{figure}

We consider a causal testing framework where inputs to an AI cause changes to its outputs. 
We adopt the conventions of {\em causal models}, a mathematical modeling technique to represent and analyze cause and 
effect relationships \cite{causality-pearl}. In a causal model, if both the input and the output have a common cause, then it forms a spurious correlation between them, and also leads to the formation of a back-door path \cite{causality-pearl}. This behavior is called {\em confounding effect} and the common cause is known as a confounder. The process of removing this confounding effect is called {\em deconfounding}. 

Now, formal definitions notwithstanding, let us consider an AI system to be fair if none of the protected attributes are affecting the output of the system. Using our causal framework, one can estimate the effect of protected attributes on the output of the system. Our approach will consist of the following main steps:
    (1) experimental apparatus consisting of data and AIs to be used, 
    (2) creating a causal framework,
    (3) performing statistical tests to assess causal dependency,
    (4) assigning ratings.

To illustrate, consider the example of Sentiment Analyzing Systems (SASs) which, as discussed earlier, have issues of bias \cite{sentiment-bias}. When given a piece of text, SASs assign a score to the text indicating whether the given text has a positive, negative or neutral sentiment (valence). Figure \ref{fig:causal-model} shows a causal model which captures the causal relations between input and output and two protected variables. The causal link from input (containing {\em Emotion Word}) to output (generating {\em Sentiment}) indicates that the emotion word affects the sentiment given by an SAS. We colored the arrow green to indicate that this causal link is desirable i.e., input should be the only attribute affecting the output. The causal links from the protected attributes to the output ({\em Sentiment}) is colored red to indicate that it is an undesirable path. If any of the protected attributes are affecting the {\em Sentiment}, then the system is said to be biased. The '?' indicates that we will be testing whether these attributes influence the sentiment of an SAS and the final rating would be based on the validity of these tests. 


Now, we prepare test inputs that utilize the different causal relationships possible between the protected variable and input (e.g., emotion words) and between themselves. The Figure shows two protected attributes, race and gender, and others can be seamlessly added. The possible values for gender could be male, female and NA, and for race could be European, African-American and NA. Here, NA means the value of the attribute is not revealed by the choice of words. Examples are: {\em He was angry} (gender = male, race = NA), {\em They are angry} (gender = NA, race = NA), {\em Amanda is angry} (gender = female, race = African-American). Here {\em angry} is an example of an emotion word and  pronouns/ (proper) nouns are taken from \cite{sentiment-bias}. 


Given statistically sufficient inputs (values of protected attributes and emotion words) and their outputs, we test the validity of each of the causal links in the proposed causal model by computing the sentiment distributions across each of the classes of the protected attributes and the deconfounded distribution of sentiment across each of the emotion words (when the protected attributes act as confounders) to give a rating to the AI system. 
A significant difference, measured using statistical tests, in the distribution across each of the classes of protected attributes would indicate presence of bias. Based on the difference between these distributions, one could assign/ implement a rating scheme (like 3-valued low, medium or high) to the system that takes into account the degree of difference found. Similarly, the difference between the deconfounded probability distribution and confounded probability distribution of sentiment across different categories of emotion words varying  due to the presence of back-door path would denote the bias present in the system. The higher the difference, the more biased the system will be. More details about the method and evaluation on a variety of SAS approaches are available in \cite{sas-rating}.



\subsection{Discussion}

Note that our causal setup is not unique to any particular type of AI (like SAS described above). In fact, it could be generalized for testing other AI  like object recognition systems \cite{mao2021generative}, recommendation systems (RS) \cite{decon-rs,causal-collab} and even estimate bias in structured datasets like  German Credit  \cite{uci-data}. But rather than aiming to remove bias in a statistical sense without understanding its social impact, we use the assessment for communicating trust information so that humans can take AI usage decisions.
Still, much needs to be done in realizing  vision of rating multimodal AI with respect to the identified research challenges so that AI could be used in real world applications in general, and trust-critical applications in particular.

\section{Conclusion}

In this paper, we presented a vision of assessing and rating AI services based on the black-box setting that they are commonly available in the marketplace, i.e., without the consumer having access to AI's code or training data. We reviewed  concerns about AI trust, 
discussed initial approaches of creating ratings for machine translators and chatbots, and laid out challenges in creating a more general approach. Then, we described a promising testing setup for AI for measuring causal dependencies between the combination of protected features and input with output, and how the dependency measurement  could be the basis of ratings. We argue that this approach would be general to handle composition of multiple data formats as well as multiple AI models that are needed for trust-critical AI applications of the future.


\bibliography{references/biplav_trust.bib,references/food.bib,references/image_explanation.bib}


\end{document}